# A Field-motion tautological approach and the role of the acceleration in setting a "quantization" condition.

**Daniel Lima Nascimento**

Institute of Physics, University of Brasilia, Brazil, daniel@fis.unb.br

**Abstract:**, In this workt is made a reanalysis of the central problem of electrodynamics, i.e., finding the conditions under which an electromagnetic field generates a stable mechanical motion and conversely the existence of this field itself can be consistent with that motion.

The fundamental assumptions of the formulation are: (i) the existence of a space-time continuum with coordinates (ct, x, y, z) that satisfy the Minkovisky metric function

$$ds^2 = c^2 dt^2 - dx^2 - dy^2 - dz^2, \qquad (1)$$

(ii) that any interaction function between particles in the continuum satisfies the requirements of covariance of the Special Theory of Relativity and (iii) that these functions should satisfy the field version of a least action principle, whose conserved quantities are given by the Noether's theorem.

The space-time metric is in fact the only real physical principle, which tells us how to connect space measurements with time intervals. Therefore, any other physical property must necessarily be derived from it through operations of tautological nature, that is, any sequence of variations will always end at the starting point. In order to produce this scheme it is necessary, however, to introduce a third element, the mass of a given object, or as is more usual in physics, a particle of mass $m$. The existence of massive objects in space-time is indeed as mysterious as the very existence of space and time [1].

We may express the interval $ds$ in Eq.(1) as a function of time, by defining the velocity vector $\mathbf{v} = (\frac{dx}{dt}, \frac{dy}{dt}, \frac{dz}{dt})$ and a dilatation factor $\gamma^{-1} = \sqrt{1-\mathbf{v}^2}$, as

$$ds = \frac{dt}{\gamma}. \qquad (2)$$

Now we introduce the motion of a particle of mass $m$ as a four-vector associated with the coordinates (t, x, y, z)[1], whose kinetic features are measured by the 3D linear momentum vector $\mathbf{p} = \gamma m \mathbf{v}$ and by the scalar kinetic energy $K$, which is obtained by multiplying Eq.(1) by $(m/ds)^2$, that is,

$$K \equiv \gamma m = \sqrt{m^2 + p^2}, \qquad (3)$$

so that, Eq.(3) stands for a measure of the particle energy of motion, whose minimum value is the particle rest mass $K_0 = m$, which occurs if and only if $p = 0$, and whose maximum value would be $K_1 = p$, for a massless "particle" (i.e., a "photon").

However, Eq.(3) has no utility if we cannot express it in terms of quantities that can be measured in the human experience context. In order to implement the tautology discussed above, we express the kinetic couple $(K, \mathbf{p})$ in terms of the "interaction" functions $(\phi, \mathbf{A})$, which are responsible by the change in the particle motion caused by an external agent, that is, the field of a proton resting at the origin, as a "force field"[2], and "internal action new coordinates" $(H, \boldsymbol{\pi})$, that are related as $H = K + \phi$, $\boldsymbol{\pi} = \mathbf{p} + \mathbf{A}$. Thus, Eq.(1) becomes

$$(\boldsymbol{\pi} - \mathbf{A})^2 - (H - \phi)^2 + m^2 = 0. \qquad (4)$$

---

[1] We use natural units now and relativistic basic notation adopted here is that of Ref. [2]

We need now to stablish equations for the unknown functions from which we can get measurable consequences. *The only way to accomplish this known up today is setting a linear differential equation for each quantity*, which could govern the evolution of it in space-time. And the only way to get a linear differential equation from a quadratic form as Eq.(4) is to construct a variational or extremum mirror of it. In order to accomplish this, we should observe that the variation of the path integral of de arc element over any finite path is identically null[3]

$$\delta \int_{s_1}^{s_2} ds = 0. \qquad (5)$$

In accordance, we shall introduce an action function S, which should satisfy a "Least Action Principle" $\delta S = 0$. Although it is in fact an extremum principle, it is historically chosen to consider it a minimum of the action function S, from where comes its name[4]. We see that $\int_{s_1}^{s_2} ds \geq 0$ because this integral would have a lesser value in another path different of the arc path.

The fundamental action component is the purely kinematic term $S = -m \int_{t_1}^{t_2} ds$, whose variation yields

$$\delta S = \int_{s_1}^{s_2} [-K \delta dt + \mathbf{p} \cdot \delta d\mathbf{r}] = -K \delta t \Big|_{s_1}^{s_2} + \mathbf{p} \cdot \delta \mathbf{r} \Big|_{s_1}^{s_2} + \int_{s_1}^{s_2} \frac{dK}{dt} dt \delta t - \int_{s_1}^{s_2} \frac{d\mathbf{p}}{dt} dt \cdot \delta \mathbf{r} = 0. \qquad (6)$$

Since both $(K, \mathbf{p})$ are constants, the third and fourth term vanishes identically. The first and second terms are required to vanish only at the extremes, but for generality and sufficiency we require for any point that $K = -\frac{\partial S}{\partial t}$ and $\mathbf{p} = \frac{\partial S}{\partial \mathbf{r}}$.

We should now make a careful distinction between the fields that act *on* the electron and those which are generated *by* it[5]. Thus the electromagnetic field of an isolated proton, as seen in Fig.1, would be completely electrostatic. Otherwise, in the presence of a moving electron, there would appear, besides the electric dipole usual field, also an electromagnetic wave, whose eikonal path, that is, the path of the source of the waves, would be the same as the electron path, as seen in Fig.2. Since these waves can transport kinetic energy away from the system, the stabilization of the motion is possible *if and only if*, some kind of stationary wave arrangement is set between the space volume containing the proton and the electron. And this implies the need of some kind of "wave mechanics" or "quantization condition", in order a collapsing motion can be avoided.

---

[2] We use this artifice to avoid by the moment the problem of the composition of two independent metrical functions.
[3] In the case of a light ray this is identically obeyed, since ds=0 along any point of the path (light interval).
[4] See for example the discussions on the subject in Ref.[3].
[5] Although Landau is our fundamental source, it must be remarked that there are these misconceptions there.

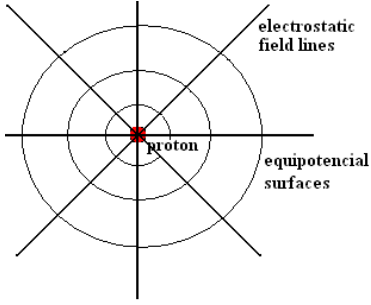 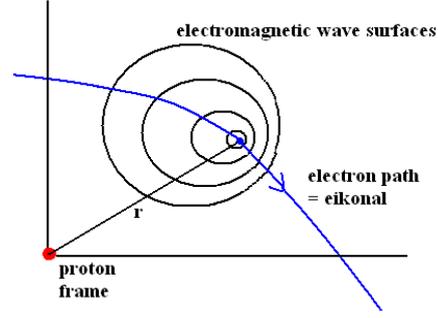

Fig.1 Electric field lines of a resting proton.  Fig.2 Electromagnetic field of a resting proton and moving electron system.

Consider now that the action function $S = S_{on} + S_{by}$ that is composed by a term that represents the action of the proton electromagnetic field *on* the electron, so that it is the cause of its motion, and another one which consider the action *by* the electron on the proton electromagnetic field, due to the electron motion, which is caused by the proton, so that the metrical tautology can be closed. They are of the form

$$S_{on} = S_{on}(\phi', \mathbf{A}', \mathbf{a}, \mathbf{v}, \mathbf{r}, t) \ , \tag{7a}$$

$$S_{by} = S_{by}(\phi, \mathbf{A}, \tfrac{\partial \phi}{\partial x}, \tfrac{\partial \phi}{\partial y}, \tfrac{\partial \phi}{\partial z}, \tfrac{\partial \phi}{\partial t}, \tfrac{\partial A_x}{\partial x}, \tfrac{\partial A_x}{\partial y}, \tfrac{\partial A_x}{\partial z}, \ldots, \tfrac{\partial A_x}{\partial t}, \tfrac{\partial A_y}{\partial t}, \tfrac{\partial A_z}{\partial t}, \mathbf{a}, \mathbf{v}, \mathbf{r}, t) \ , \tag{7b}$$

where the primes are to distinguish among *on* and *by* forms of the field potentials. For $S_{on}$ the field "coordinates" are considered constant and for $S_{by}$, which depends also on the first order derivatives of the field potentials, the particle velocity and acceleration should be considered constant in the variation process.

Thus the action function on the electron is

$$S_{on} = \int_{t_1}^{t_2}(-m\,ds + \mathbf{A}'\cdot d\mathbf{r} + \mathbf{W}\cdot d\mathbf{v} - \phi'\,dt) = \int_{t_1}^{t_2}(-m\sqrt{1-\mathbf{v}^2} + \mathbf{A}'\cdot\mathbf{v} + \mathbf{W}\cdot\mathbf{a} - \phi')\,dt \equiv \int_{t_1}^{t_2} L_{on}\,dt \ . \tag{8}$$

First of all, we need to generalize the relation between action and momentum, i.e.,

$$H = -\frac{\partial S}{\partial t}, \qquad \boldsymbol{\pi} = \frac{\partial S}{\partial \mathbf{r}}. \tag{9}$$

Secondly, we consider the variational problem on the Lagrangean function $L_{on}$ in which we consider variations of the motion degrees of freedom, keeping the field coordinates constant:

$$\delta \int_{t_1}^{t_2} L_{on}\,dt = 0. \tag{10}$$

Proceeding to the variations we get, by using integration by parts,

$$\int_{t_1}^{t_2}\left(\frac{\partial L_{on}}{\partial \mathbf{r}}\delta\mathbf{r} + \frac{\partial L_{on}}{\partial \mathbf{v}}\delta\mathbf{v} + \frac{\partial L_{on}}{\partial \mathbf{a}}\delta\mathbf{a}\right)dt =$$

$$\int_{t_1}^{t_2}\left[\frac{\partial L_{on}}{\partial \mathbf{r}} - \frac{d}{dt}\frac{\partial L_{on}}{\partial \mathbf{v}} + \frac{d^2}{dt^2}\left(\frac{\partial L_{on}}{\partial \mathbf{a}}\right)\right]\delta\mathbf{r}\,dt + \left(\frac{\partial L_{on}}{\partial \mathbf{v}} - \frac{d}{dt}\frac{\partial L_{on}}{\partial \mathbf{a}}\right)\delta\mathbf{r}\Big|_{t_1}^{t_2} + \frac{\partial L_{on}}{\partial \mathbf{a}}\delta\mathbf{v}\Big|_{t_1}^{t_2} = 0 \tag{11}$$

and considering that $\delta\mathbf{r}(t_1) = \delta\mathbf{r}(t_2) = 0$ and $\delta\mathbf{v}(t_1) = \delta\mathbf{v}(t_2) = 0$, we arrive at

$$\frac{d^2}{dt^2}\left(\frac{\partial L_{on}}{\partial \mathbf{a}}\right) - \frac{d}{dt}\left(\frac{\partial L_{on}}{\partial \mathbf{v}}\right) + \frac{\partial L_{on}}{\partial \mathbf{r}} = 0. \tag{12}$$

Considering now the Lagrangean in Eq.(8)

$$L_{on} = -m\sqrt{1-\mathbf{v}^2} + \mathbf{A}' \cdot \mathbf{v} - \phi' + \mathbf{W} \cdot \mathbf{a}, \tag{13}$$

with $\mathbf{A} = \mathbf{A}(\mathbf{r},t)$ and $\mathbf{W}(\mathbf{v},\mathbf{r},t)$, we can see immediately that

$$\boldsymbol{\pi} = \frac{\partial L}{\partial \mathbf{v}} = \mathbf{p} + \mathbf{A}' + (\mathbf{a} \cdot \tfrac{\partial}{\partial \mathbf{v}})\ \mathbf{W} + \mathbf{a} \times (\tfrac{\partial}{\partial \mathbf{v}} \times \mathbf{W}) \tag{14}$$

and, calculating the Lagrangean derivatives in Eq(12), we get the following equation of motion

$$\frac{d\mathbf{p}}{dt} = \mathbf{E}' + \mathbf{v} \times \mathbf{B}' + \mathbf{R}, \tag{15}$$

where the first two field vectors are the electric and magnetic vectors respectively, yet the last one is related to retardation or dissipation effects due to the acceleration:

$$\mathbf{E}' = -\frac{\partial \mathbf{A}'}{\partial t} - \frac{\partial \phi'}{\partial \mathbf{r}}, \quad \mathbf{B}' = \tfrac{\partial}{\partial \mathbf{r}} \times \mathbf{A}', \tag{16}$$

$$\mathbf{R} = \mathbf{a} \times (\tfrac{\partial}{\partial \mathbf{r}} \times \mathbf{W}) - (\mathbf{a} \cdot \tfrac{\partial}{\partial \mathbf{r}})\ \mathbf{W} + \frac{d^2 \mathbf{W}}{dt^2} - \frac{d}{dt}\left[ (\mathbf{a} \cdot \tfrac{\partial}{\partial \mathbf{v}})\ \mathbf{W} + \mathbf{a} \times (\tfrac{\partial}{\partial \mathbf{v}} \times \mathbf{W}) \right]. \tag{17}$$

Thus, Eq(15) gives the motion of the particle under the given $\mathbf{E}'$, $\mathbf{B}'$ and $\mathbf{R}$. It can immediately be seen that Eqs.(16) correspond to the "inner" Maxwell equations,

$$\tfrac{\partial}{\partial \mathbf{r}} \times \mathbf{E}' = -\partial \mathbf{B}'/\partial t, \qquad \tfrac{\partial}{\partial \mathbf{r}} \cdot \mathbf{B}' = 0. \tag{18}$$

We now try the converse, that is, for a given motion ($\mathbf{a},\mathbf{v},\mathbf{r}$), we shall get the corresponding equations for the fields $\mathbf{E} \neq \mathbf{E}'$ and $\mathbf{B} \neq \mathbf{B}'$, which are, in general, not the same as given in Eqs.(16), which are generated by that motion, so that the field-motion tautology is near to be closed. To accomplish this, we must get $S_{by}$, by transforming the path-dependent form of the Lagrangean function to a volume dependent one, since the fields are vector functions that can propagate through the whole space for any time. As a consequence, it must also be supposed that the particle charge, which is contained in its mass, is distributed over the space while it remains being a point particle. The "solution" for this inconsistence is obtained by using the idealized density function created by Dirac, which is zero anywhere, with exception to the particle position, where it is infinity. If the particle has a finite size, all the calculations to be made below would have to be reviewed. Thus $\rho(\mathbf{r},\mathbf{r}') = \delta(\mathbf{r}-\mathbf{r}')$ and $\int_{space} \rho dV = 1$. By using this property and also the definitions $\mathbf{E} = -\frac{\partial \mathbf{A}}{\partial t} - \frac{\partial \phi}{\partial \mathbf{r}}$, $\mathbf{B} = \tfrac{\partial}{\partial \mathbf{r}} \times \mathbf{A}$, we get the field action function and corresponding Lagrangean density as

$$S_{by} = \int_{t_1}^{t_2} dt \int_{space} dV \left[ \mathbf{A} \cdot \rho \mathbf{v} - \rho \phi + \mathbf{W} \cdot \rho \mathbf{a} + \tfrac{1}{2}(\mathbf{E}^2 - \mathbf{B}^2) \right] \equiv \int_{t_1}^{t_2} dt \int_{space} dV\ L_{by}. \tag{19}$$

where the three terms in which $\rho$ appears are responsible by the interaction between the proton field and the electron path and the last term is responsible by the existence of the free field, that is, electromagnetic waves with a straight line eikonal generated by a resting source. The variation problem $\delta S_{by} = 0$ becomes now

$$\int_{t_1}^{t_2}\int_{space}\left[\frac{\partial L_{by}}{\partial \frac{\partial \phi}{\partial \mathbf{r}}}\cdot\delta\frac{\partial \phi}{\partial \mathbf{r}}+\frac{\partial L_{by}}{\partial \frac{\partial \phi}{\partial t}}\delta\frac{\partial \phi}{\partial t}+\frac{\partial L_{by}}{\partial \frac{\partial A_x}{\partial \mathbf{r}}}\cdot\delta\frac{\partial A_x}{\partial \mathbf{r}}+\frac{\partial L_{by}}{\partial \frac{\partial A_y}{\partial \mathbf{r}}}\cdot\delta\frac{\partial A_y}{\partial \mathbf{r}}+\frac{\partial L_{by}}{\partial \frac{\partial A_z}{\partial \mathbf{r}}}\cdot\delta\frac{\partial A_z}{\partial \mathbf{r}}+\right.$$
$$\left.\frac{\partial L_{by}}{\partial \frac{\partial A_x}{\partial t}}\delta\frac{\partial A_x}{\partial t}+\frac{\partial L_{by}}{\partial \frac{\partial A_y}{\partial t}}\delta\frac{\partial A_y}{\partial t}+\frac{\partial L_{by}}{\partial \frac{\partial A_z}{\partial t}}\delta\frac{\partial A_z}{\partial t}\right]dVdt=0, \tag{20}$$

which, by using integration by parts, the Gauss theorem and supposing that the potential variations vanish at the bound surface at the infinity, $\delta\phi|_{S_\infty}=\delta\mathbf{A}|_{S_\infty}=0$, yields

$$\frac{\partial}{\partial \mathbf{r}}\cdot\frac{\partial L_{by}}{\partial \frac{\partial \phi}{\partial \mathbf{r}}}+\frac{\partial}{\partial t}\frac{\partial L_{by}}{\partial \frac{\partial \phi}{\partial t}}-\frac{\partial L_{by}}{\partial \phi}=0, \tag{21a}$$

$$\frac{\partial}{\partial \mathbf{r}}\cdot\frac{\partial L_{by}}{\partial \frac{\partial A_a}{\partial \mathbf{r}}}+\frac{\partial}{\partial t}\frac{\partial L_{by}}{\partial \frac{\partial A_a}{\partial t}}-\frac{\partial L_{by}}{\partial A_a}=0, \tag{21b}$$

where in the later term $a=x,y,z$ for the components of the vector potential.

Using the Lagrangean density of Eq. (19)

$$L_{by}=\mathbf{A}\cdot\rho\mathbf{v}-\rho\phi+\mathbf{W}\cdot\rho\mathbf{a}+\tfrac{1}{2}\ (\mathbf{E}^2-\mathbf{B}^2)\ , \tag{22}$$

we arrive at the second pair of the Maxwell equations, as expected[6],

$$\tfrac{\partial}{\partial \mathbf{r}}\times\mathbf{B}=\partial\mathbf{E}/\partial t+\rho\mathbf{v},\qquad \tfrac{\partial}{\partial \mathbf{r}}\cdot\mathbf{E}=\rho. \tag{23}$$

Thus, it is possible, in principle, to close the tautology field-motion[7].

We can also obtain independent wave equations for the field potentials by substituting their definitions into Eq(23) and using some vector identities

$$\Delta\mathbf{A}-\frac{\partial^2\mathbf{A}}{\partial t^2}=-\rho\mathbf{v},\qquad \Delta\phi-\frac{\partial^2\phi}{\partial t^2}=-\rho, \tag{24}$$

in which it was necessary to make use of the usual Lorentz gauge $\tfrac{\partial}{\partial \mathbf{r}}\cdot\mathbf{A}+\partial\phi/\partial t=0$, because of the high degree of arbitrariness from what they arise.

We can take a first look at the possibility of a mutually stable field-motion configuration into which the kinetic energy could stay unchanged in average, under the action of the field of force. Then, by the motion side we see from Eq.(15) that

$$\frac{dK}{dt}=\mathbf{v}\cdot\frac{d\mathbf{p}}{dt}=\mathbf{v}\cdot\mathbf{E}'+\mathbf{v}\cdot\mathbf{R}. \tag{25}$$

Now, multiplying each time derivative in the Maxwell equations, Eq(23), by the respective field, and recalling that $\tfrac{\partial}{\partial \mathbf{r}}\times\mathbf{E}=-\partial\mathbf{B}/\partial t$ and $\tfrac{\partial}{\partial \mathbf{r}}\cdot\mathbf{B}=0$, we get

$$\mathbf{E}\cdot\partial\mathbf{E}/\partial t=\mathbf{E}\cdot\tfrac{\partial}{\partial \mathbf{r}}\times\mathbf{B}-\rho\ (\mathbf{E}-\mathbf{E}')\ \cdot\mathbf{v}-\rho\mathbf{E}'\cdot\mathbf{v}, \tag{26a}$$

$$\mathbf{B}\cdot\partial\mathbf{B}/\partial t=-\mathbf{B}\cdot\tfrac{\partial}{\partial \mathbf{r}}\times\mathbf{E}, \tag{26b}$$

which once summed and volume integrated and using the Gauss theorem finally yields

$$\frac{d}{dt}\left[K+\tfrac{1}{2}\int_{space}dV\ (\mathbf{E}^2+\mathbf{B}^2)\ \right]=-\int_{surface}d\mathbf{F}\cdot\ (\mathbf{E}\times\mathbf{B})\ +\mathbf{v}\cdot(\ \mathbf{R}+\mathbf{E}'-\mathbf{E}) \ . \tag{27}$$

---

[6] This was possible only by supposing that $\mathbf{W}=\mathbf{W}(\mathbf{v},\mathbf{r},t)$, that is, it does not depend on $\phi,\mathbf{A}$.

[7] It must be observed that we have not yet made any hypothesis about the electron charge or field strength.

We see at once that the tuning field condition $\mathbf{E}' = \mathbf{E}$ is quite complicated, because we should recall that $\mathbf{E}(\mathbf{r}',t)$ is a high frequency radiation field caused by the electron motion that acts on another object, at $\mathbf{r}',t$, outside the electron-proton system, while $\mathbf{E}'(\mathbf{r},t)$ is the low frequency proton electric field that acts on the electron, at $(\mathbf{r},t)$, which is in fact a limiting case of $\mathbf{E}(\mathbf{r}',t)$ when $\mathbf{r}' \to \mathbf{r}$, that is, it would be a self-effect. Therefore, any consideration on the possibility of a steady motion configuration for the system depends on which kind of resonance between on and by fields is possible to be reached. Some possibilities of handeling the motion-field interaction are discussed below, but this problem remains in its deepest grounds stills an open problem.

**The classical atom model and its "unavoidable" collapse.**

Within the context of the previous general approach, this is in fact an illusory problem, since field and motion are in mutual attempt to conform to each other. Anyway, the collapse would be unavoidable only on the exceptional conditions in which $\mathbf{E}' = \mathbf{E}$ and no reaction term were present. In this case, the motion-Lagrangean would be non dependent on the acceleration, thus $\mathbf{R} = 0$, and there would be no way of compensating the loss of energy due to the Poynting flux integral. Therefore the motion would collapse and the electron would free "fall" toward the center of force, its energy tending to minus infinity [2].

**Sommerfeld quantization condition.**

If acceleration dependence is allowed, the motion can be stabilized by obtaining field configurations that can be able to restore any tendency to collapse of the system. This possibility was first tried by A. Sommerfeld, in 1916 [4], in the context of the "old" quantum theory. It consisted in stabilizing directly the generalized linear momentum in Eq.(14) through an action integral, which is imposed to be an integer

$$\oint \boldsymbol{\pi} \cdot d\mathbf{r} = \oint \left( \mathbf{p} + \mathbf{A} \right) \cdot d\mathbf{r} + \oint \left[ (\mathbf{a} \cdot \tfrac{\partial}{\partial \mathbf{v}}) \mathbf{W} + \mathbf{a} \times (\tfrac{\partial}{\partial \mathbf{v}} \times \mathbf{W}) \right] \cdot d\mathbf{r} = n, \tag{28}$$

in which is usual the further assumption $\mathbf{A} = 0$. It may be separated in two independent conditions

$$\oint \mathbf{p} \cdot d\mathbf{r} = n, \qquad \oint \left[ (\mathbf{a} \cdot \tfrac{\partial}{\partial \mathbf{v}}) \mathbf{W} + \mathbf{a} \times (\tfrac{\partial}{\partial \mathbf{v}} \times \mathbf{W}) \right] \cdot d\mathbf{r} = 0, \tag{29}$$

the first line integral corresponding to the Sommerfeld quantization condition, and the second would be a consequence of the stationary configuration on the fields, although this has never been made explicitly.

**The quantization condition of the Sthochastic Electrodynamics.**

A second possibility to stabilize the motion is that adopted by the so called "Sthocastic Eletrodynamics" [5-6]. In this formulation there is no acceleration dependence in the motion Lagrangean, so that the reaction field would vanish, $\mathbf{R} = 0$. It is then postulated a metaphysical zero point field out of which there appears a compensating field, that takes the place of the reaction field $\mathbf{R}$ in the exact proportion of the loss of energy due to the Poynting flux, that is,

$$\int_{surface} d\mathbf{F} \cdot (\mathbf{E} \times \mathbf{B}) + \mathbf{v} \cdot \mathbf{R} = 0 \ . \tag{30}$$

Thus, the system would reach the field-motion stable configuration

$$K + \tfrac{1}{2} \int_{space} dV \ (\mathbf{E}^2 + \mathbf{B}^2) = E = \text{constant} \ . \tag{31}$$

This would be a very good result, it were not the metaphysical character of the zero point field behind it.

A third possibility is that followed directly or indirectly by all the current formulations of quantum mechanics. It consists in imposing a variational condition that stabilizes the Hamiton-Jacobi Eq.(4) directly, through the use of a regularized action function or "wave function" and using an electrostatic potential. The first attempt was made in a non relativistic context by Schroedinger in 1926 [7], followed by a non succeeded relativistic attempt by Klein [8] and Gordon [9] and by the succeeded formulation by Dirac in 1928 [10].

As a fourth possibility, we have reconsidered the Schroedinger formulation and constructed out of it a formulation in which the electron can follow a classical two-dimensional path until a stable attractor is reached [11]. We also have made an approach alternative to Dirac´s formulation in which, by using classical conservation theorems before the Dirac matrices are found, we have found a 2D form of the Dirac equation whose two-spinor solutions are very easy to get [12], allows to follow an electron path and makes possible to obtain an extremely accurate numerical method to find its energy eigenvalues [13].

The two particle problem, which demands a two metric functions simultaneous solution, is already in preparation, allowing higher order energy corrections for the hydrogen atom, as well as a solution for the helium atom [14,15] which will be shown to be very simple with respect to the Breit formulation of 1929 [16].

In conclusion, we have pointed out that the fundamental electrodynamical problem has in no way found a completely consistent solution in its two independent aspects: field and motion. So that any existing formulation is still precarious and a definite solution still remains to be found.

## References.